\documentclass{iopart}
\pdfoutput=1
\usepackage{iopams}
\usepackage{setstack}
\usepackage{graphicx}
\usepackage{epsfig}
\usepackage{color}

\begin{document}
\title[LSC Glitch Group]{The LSC Glitch Group : Monitoring Noise Transients during the fifth LIGO Science Run}
\author{L~Blackburn$^{1}$, L~Cadonati$^{2}$, S~Caride$^{3}$, S~Caudill$^{4}$, S~Chatterji$^{5}$, N~Christensen$^{3}$, J~Dalrymple$^{6}$, S~Desai$^{7}$, A~Di~Credico$^{6}$, G~Ely$^{3}$, J~Garofoli$^{8}$, L~Goggin$^{5}$, G~Gonz\'alez$^{4}$, R~Gouaty$^{4}$,  C~Gray$^{8}$, A~Gretarsson$^{9}$, D~Hoak$^{10}$, T~Isogai$^{3}$, E~Katsavounidis$^{1}$, J~Kissel$^{4}$, S~Klimenko$^{11}$, R A~Mercer$^{11}$, S~Mohapatra$^{2}$, S~Mukherjee$^{12}$, F~Raab$^{8}$, K~Riles$^{13}$, P~Saulson$^{6}$, R~Schofield$^{14}$, P~Shawhan$^{15}$, J~Slutsky$^{4}$, J~R~Smith$^{6}$, R~Stone$^{12}$, C~Vorvick$^{8}$, M~Zanolin$^{9}$, N~Zotov$^{16}$ and  J~Zweizig$^{5}$}

$^{1}$LIGO-Massachusetts Institute of Technology, 
Cambridge, MA 02139, USA

$^{2}$ University of Massachusetts, Amherst, MA 01003, USA  

$^{3}$ Carleton College, Northfield, MN 55057, USA

$^{4}$ Louisiana State University, Baton Rouge, LA 70803, USA

$^{5}$ LIGO - California Institute of Technology, 
Pasadena, CA 91125, USA

$^{6}$ Syracuse University, Syracuse, NY 13244, USA

$^{7}$ The Pennsylvania State University, University Park, PA 16802, USA

$^{8}$ LIGO Hanford Observatory, Richland, WA 99352, USA

$^{9}$ Embry-Riddle Aeronautical University, Prescott, AZ 86301, USA

$^{10}$ LIGO Livingston Observatory, Livingston, LA 70754, USA

$^{11}$ University of Florida, Gainsville, FL 32611, USA 

$^{12}$ The University of Texas, Brownsville, TX 78520, USA

$^{13}$ University of Michigan, Ann Arbor, MI 48109, USA

$^{14}$ University of Oregon, Eugene, OR 97403, USA

$^{15}$ University of Maryland, College Park, MD 20742, USA

$^{16}$ Louisiana Tech University, Ruston, LA 71272, USA

\ead{desai@gravity.psu.edu} 
\begin{abstract}
The LIGO Scientific Collaboration (LSC) {\em glitch group} is part of the LIGO detector characterization
effort. It consists of data analysts and detector experts who, during and after science runs, collaborate for 
a better understanding of noise transients in the detectors. Goals of the glitch group during the fifth LIGO
science run (S5) included (1) offline  assessment of the detector data quality,
with focus on noise transients, (2)  veto recommendations for astrophysical analysis  and (3) feedback to the 
commissioning team on anomalies seen
in gravitational wave and auxiliary data channels. 
Other activities included the study of auto-correlation of triggers from burst searches, stationarity of the detector noise and veto
studies. The group identified causes for several noise transients that triggered false alarms in the gravitational wave searches; the times of such transients were identified and vetoed from the data generating 
the LSC astrophysical results.  

\end{abstract}
\maketitle

\section {Introduction}

The ``glitch group'' is one of the subgroups of the Detector Characterization
Committee within the LIGO Scientific Collaboration (LSC). 
In this
paper we shall use the term  ``glitch'' to denote any short-duration noise 
transient in the gravitational wave
channel as well as transients in auxiliary channels. Glitches 
produced by environmental effects or instrumental malfunctions are a source of background for 
transient gravitational wave signals, such as unmodelled bursts or compact binary coalescences.
Sufficiently strong glitches are also responsible for loss of lock and
decreased observation time.
The glitch group was established  in 2003 to characterize noise transients in LIGO. At times  these investigations revealed
causes that could be fixed and those transients were eliminated.
The group  consists of  members from the  analysis groups searching for short duration gravitational waves from 
coalescing binary systems, supernovae, or other astrophysical systems,  as well as detector experts and
operators from both the LIGO sites at Hanford and Livingston. There is 
substantial interaction between the glitch group and other detector characterization
working groups such as {\sf Calibration}, {\sf Data Quality}, {\sf Dataset Reduction}, 
{\sf Environmental Disturbances}, and {\sf Hardware Injections}. This article will focus on the activities and findings of the 
glitch group during the fifth LIGO science run (``S5'') which  started in November 2005 and 
ended in  October 2007~\cite{LIGOdet}. In Sect.~\ref{sec:methods} we describe the tools used by the group for the diagnoses of instrument artifacts. In Sect.~\ref{sec:results} we describe some of the findings that led to  elimination of identified problems and in better data quality.
Finally, we shall conclude by highlighting some of the 
post-S5 activities (Sect.~\ref{sec:posts5}).

The glitch group provided a forum for experts in data analysis and detector commissioning to join forces, brainstorm 
and  assess the performance of the  LIGO detectors during the S5 run (4th November 2005 - 1st October 2007). 

Goals of the glitch group during the fifth LIGO
science run (S5) included (1) offline  assessment of the detector data quality,
with focus on noise transients, (2)  veto recommendations for astrophysical analysis  and (3) feedback to the 
commissioning team on anomalies seen in gravitational wave and auxiliary data channels. 

The glitch group activities complemented  
realtime  investigations and onsite detector troubleshooting,  and provided guidance to the burst and compact binary 
coalescence (CBC) analysis groups in their veto choices.

 Members of the glitch group conducted offsite  shifts, 
each covering 3-4 days of data acquisition. Results from these shifts were discussed in weekly 
telephone conferences.  
Highlights from these shifts were also presented each week
in the run coordination and detector characterization teleconferences. Detailed specialized investigations were carried out by individual glitch group members.

\section{Goals and Methods}
\label{sec:methods}
Glitch group members analyzed LIGO data with several near real-time 
algorithms, with latency
 ranging from  few minutes  to a day. Some of these were run using the 
Data Monitoring Tool ({\tt DMT}) environment~\cite{dmt}
within LIGO. The {\tt DMT} is a set of algorithms that monitor various  aspects
of LIGO data quality, display status information and record data quality statistics. The goals of these 
near online algorithms  ranged from searches 
for gravitational wave signals from unmodelled bursts and inspirals to studies of detector noise.

{\tt Block-Normal} : {\tt Block-Normal} is an algorithm designed to search for 
short-duration unmodelled gravitational wave bursts.  It is based on a time domain analysis
of the  data and uses a Bayesian statistics figure of merit to select candidate events~\cite{bn}.
During a glitch shift, we scanned single interferometer outliers from {\tt Block-Normal}
 using the event visualization tools.

\begin{figure}
\begin{center}
  \includegraphics[width=0.5\textwidth]{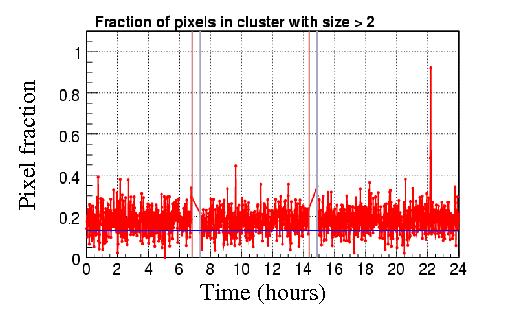}
  \caption{\label{burstmonfig} 
{\tt BurstMon}~\cite{burstmon} pixel fraction as a function of time (with one entry per minute).  
{\tt BurstMon} first identifies the most significant 10 percent of 
time-frequency "pixels" using a wavelet based decomposition of the data.  
The pixel fraction is defined as the fraction of these significant                                            
pixels that can be grouped into clusters of two or more pixels.                                           
Ideal noise typically produces isolated pixels, resulting in a                                            
typical pixel fraction of 0.13.  Pixel fractions much greater                                             
than this value indicate a high rate of glitches in the data,                                             
which preferentially produce clusters of significant pixels.}

\end{center}
\end{figure}

{\tt BurstMon} :
{\tt BurstMon} is the DMT tool for monitoring the burst detection performance  of LIGO detectors. It is closely 
related to the {\tt Waveburst} algorithm~\cite{wb} that is used for untriggered gravitational wave burst searches 
using data from second, third and fourth LIGO science runs. 
This monitor produces 3 figures of merits: a measure of the rate of non-stationarity 
called ``pixel fraction'', the real-time detector sensitivity to gravitational wave bursts, and noise 
variability in various frequency bands. A sample plot from the ``pixel fraction''
figure of merit  is shown in Fig.~\ref{burstmonfig}.
The term ``pixel'' is used to denote a time-frequency bin and the pixel fraction is  the 
fraction of pixels which can be grouped into clusters of two or more pixels.
It indicates what  fraction of the time-frequency volume is affected by the non-Gaussianity of the detector noise.
For stationary Gaussian noise, its value is equal to 0.13. However for real LIGO data it could be as large as 1.
 More details on  the {\tt Burstmon} figures of merit are provided in 
Ref.~\cite{burstmon}.

 {\tt InspiralMon} : 
Online searches for inspiralling binary compact objects were done using matched-filter based searches for 
compact object mergers between 1 and 3 $M_{\odot}$ using second order post-Newtonian 
stationary phase templates~\cite{inspiral}. All separate triggers within a 15~s time-window were 
clustered into one set. These triggers were not used for the actual gravitational wave search (which uses coincidence between detectors, a larger template bank and several signal-based vetoes), 
but were very useful for diagnostic purposes.
The signal to noise ratio of the loudest trigger was  displayed in the control room every minute. During the glitch 
shifts, we examined the loudest 20 single interferometer CBC triggers produced each day 
with  signal-to-noise ratio greater than 15.

\begin{figure}
\begin{center}
  \includegraphics[width=0.5\textwidth]{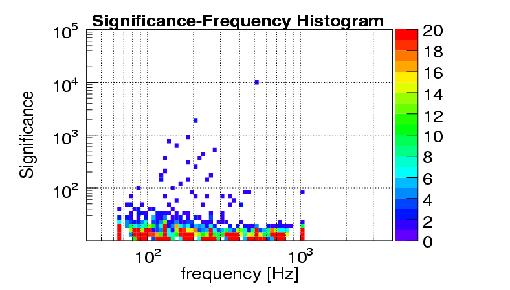}
  \caption{\label{sigvsfreq} {\tt KleineWelle} significance versus  frequency (over a one-day 
integrated period).
The color-scale is the number of events in a given significance-frequency bin.}
\end{center}
\end{figure}

\begin{figure}
\begin{center}
\includegraphics[width=0.5\textwidth]{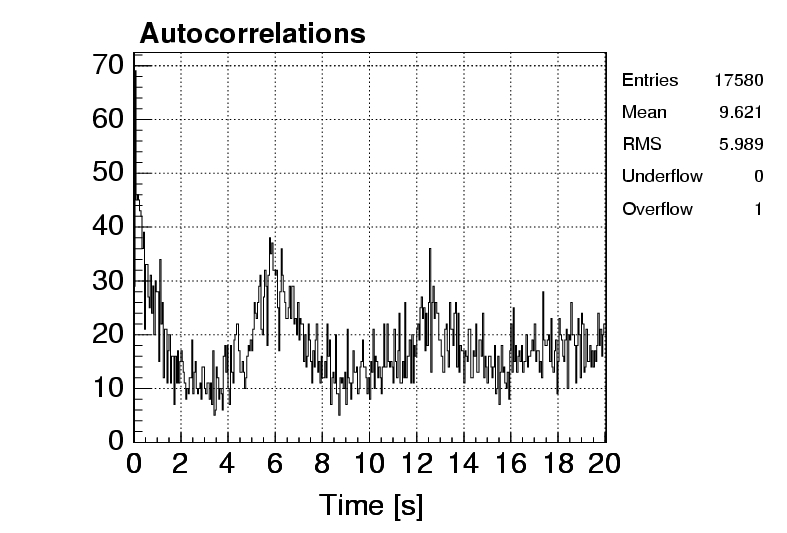}
\caption{\label{autocorr} Auto-correlogram of the {\tt KleineWelle} triggers. The auto-correlogram is a 
histogram of the time-difference between a given trigger and all other triggers. A peak in this plot is an 
indicator of periodicity which on a time-scale of a few seconds is due to enhanced microseismic noise.}
\end{center}
\end{figure}

 {\tt KleineWelle} :
{\tt KleineWelle}~\cite{kw} is  a  single interferometer  event trigger generator. It is based on 
the dyadic wavelet decomposition of a time-series. The wavelet transform provides time-frequency 
localization of signal energy represented by the wavelet coefficients of the decomposition. During S5, 
{\tt KleineWelle} analyzed in near-realtime (and offline) the gravitational wave channel and  a variety of auxiliary 
channels for the three LIGO detectors and GEO. A variety of diagnostic plots were produced from
these triggers. Multi-dimensional classification analysis was also done using these triggers~\cite{multidim}.
Plots of the the trigger 
rate for a given channel with low and high thresholds could be produced with a graphical  
web-based interface.  
During the glitch shifts, we explored
both double-coincident (between the 2 LIGO Hanford interferometers) and triple-coincident {\tt KleineWelle} triggers
using event visualization tools. We also examined various other diagnostic plots such as 
trigger auto-correlations, trigger periodicities, trigger significance as a function of frequency etc.  
Anomalous features in the auto-correlation plots are 
usually due to enhanced microseismic noise. Some of these plots are shown in 
Figs.~\ref{sigvsfreq} and ~\ref{autocorr}.

{\tt NoiseFloorMon} : 
{\tt NoiseFloorMon} is a monitor to detect slow drifts in the noise floor~\cite{noisefloormon}.
It was applied to the gravitational wave channel and to various seismic channels.
During glitch shifts we typically looked at minute trends of threshold crossings and cross-correlations
with seismic channels.

\begin{figure}
\begin{center}
  \includegraphics[width=0.5\textwidth]{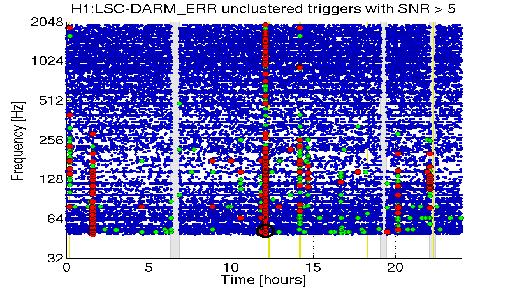}
  \caption{\label{qonlinefig} Time-frequency scatter plot of triggers from the {\tt QOnline} analysis of         
 the 4 km LIGO Hanford detector. The color indicates ranges of SNR                                        
 with blue indicating SNRs from 5 to 10, green from 10 to 20, and red                                      
 greater than 20.}
\end{center}
\end{figure}

{\tt QOnline} : The {\tt QOnline} pipeline  is an online  multi-resolution time-frequency search for 
statistically significant excess signal energy. It is equivalent to a templated matched filter search (in 
the whitened signal space),  
whose basis functions are sinusoidal Gaussians of varying central time, central frequency, 
and the quality factor $Q$. Details  on the Q-transform are provided in Ref.~\cite{shourovthesis}.
The algorithm  was run online on data from  the three LIGO detectors, VIRGO and GEO. During the glitch shifts, we examined
  the trigger trends  from the
{\tt QOnline} pipeline (See Fig.~\ref{qonlinefig}) and a scan of the loudest event within each hour.

\begin{figure}
\begin{center}
 \includegraphics[width=0.5\textwidth]{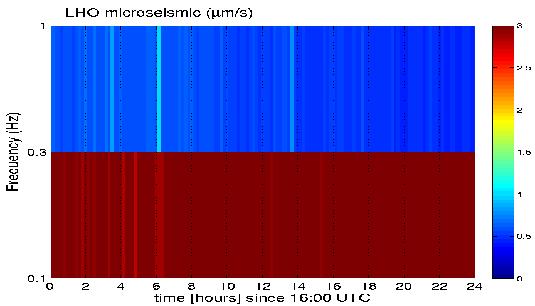}
  \caption{\label{museism} Band-limited seismic noise in the 0.1 - 1 Hz band at Hanford. Such high microseism
could cause peaks in the auto-correlogram seen in Fig.~\ref{autocorr}.}
\end{center}
\end{figure}

Besides the above near-online analysis, we also studied   online 
figures of merit (which are usually produced in realtime in the control room) such as 
the effective distance to which LIGO is sensitive to binary neutron star inspirals, as well as environmental
factors like wind, band-limited seismic noise, etc. See Fig.~\ref{museism} for a plot of band-limited 
microseismic noise
usually looked at during these shifts.



We also gained understanding of glitch mechanisms by listening
to the whitened versions of glitch waveforms through high-quality audio
systems, taking advantage of the fact that our search is carried
out at audio frequencies. A wide range of input disturbances leads to
glitches with no discernible differences. In fact, each interferometer
seems to have a characteristic glitch waveform, each a variation of a
few-cycle oscillation near 100 Hz. Further study is exploring the
exceptions to this general rule, including longer-duration (``more
musical'') tones, broad-band glitches, and echoes. It is hoped that
these studies will give a clue about the glitch mechanism(s), still
undiagnosed.

\section{Event visualization tools}
We used two event visualization tools for a  better 
insight into 
the behavior of detectors at any particular time of interest, that provide  snapshots 
of the  LIGO auxiliary and environmental channels as well as  the gravitational wave channel. 
These are similar to event display tools routinely used in high energy physics experiments
to depict the tracks of particles.
Times of interest  included outliers from the burst and
CBC searches, hardware signal injections, gamma-ray-burst arrival 
times, environmental injections, etc. The tools provided  insight into the behavior of the detectors 
at a given time and helped identify a few ``smoking  gun'' causes of loud glitches, data corruption and sources of  
lock-loss.

\begin{figure}
\begin{center}

 \includegraphics[width=0.5\textwidth]{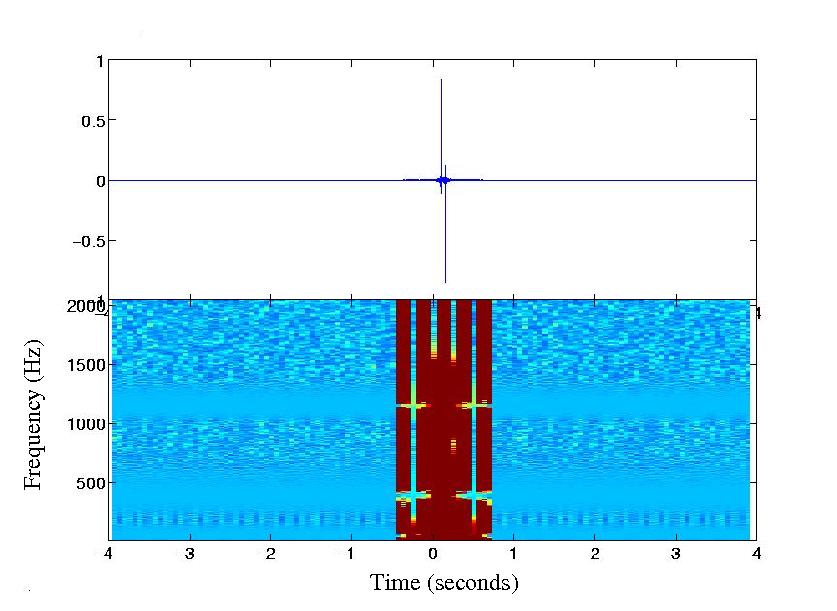}
  \caption{ \label{edspecgram} Superposed time-series and median-normalized spectrogram as it appears in the {\tt Event-Display} of a glitch in the calibration channel, after the application of linear predictive error filter (defined in Ref.~\cite{shourovthesis}).}
\end{center}
\end{figure}

{\tt Event-Display} :
The {\tt Event-Display} is also a  web-based event visualization tool which shows the   time-series 
and frequency spectrograms of a fixed set of channels along with various diagnostic information on the 
state of the detectors at that time, and output from the {\tt Parameter Estimation}~\cite{Sylvestre02} code. 
The intensity in a given time-frequency bin is normalized by the median.
One example of this type of  specialized spectrogram of a glitch in the calibration channel is shown in Fig.~\ref{edspecgram}.

\begin{figure}
\begin{center}
  \includegraphics[width=0.5\textwidth]{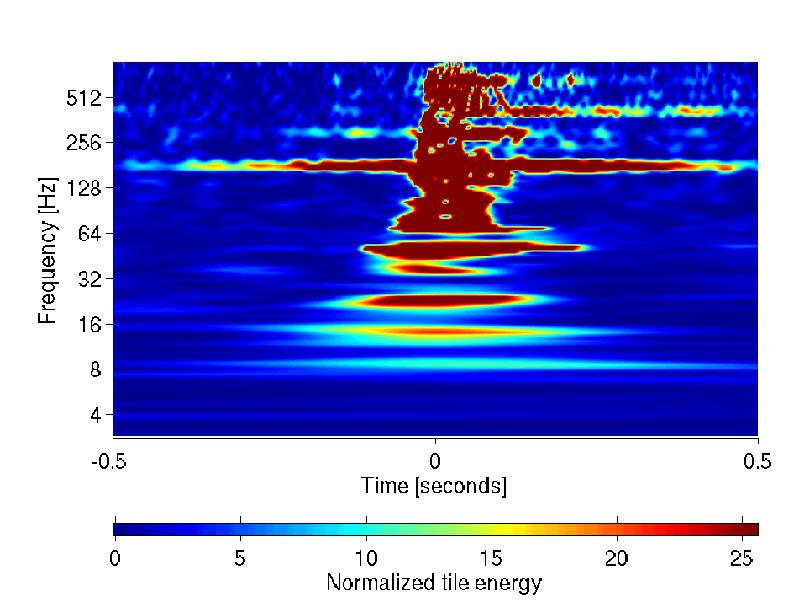}
  \caption{\label{qscanplot} {\tt QScan} of a glitch in 
a voltmeter channel, after data whitening with a
linear predictive error filter (defined in Ref.~\cite{shourovthesis}).}
\end{center}
\end{figure}

{\tt QScan} : 
{\tt QScan} is used to investigate multiple detector channels 
around times of interest. {\tt QScan} produces ``Q spectrogram'' displays, based on the same transform
used by the {\tt QOnline} analysis. For statistically significant channels, {\tt QScan} 
produces thumbnails of the time-series and ``Q spectrograms'' in 3 different time-windows ($\pm$ 0.5 sec., $\pm$ 2 sec., and $\pm$ 8 sec.)
 on a webpage.  The list of channels to look at can be defined with a  configuration file.
This tool has been extensively used in the control room by operators and science monitors
to diagnose lock-losses and is also used to look at the  various channels in the VIRGO detector. A {\tt QScan} of a glitch
in the voltmeter channel is shown  in Fig.~\ref{qscanplot}.

\section{Some results from the glitch group efforts}
\label{sec:results}
During S5, the glitch group provided an offline forum to explore and discuss the day-to-day performance 
of the LIGO detectors, and provide commissioners  with valuable feedback from a data analysis perspective. 
More importantly, these investigations   led to the 
creation of several data quality flags which are used as vetoes in the analysis 
of  S5 data.  The identification of noise transients is particularly important for the development 
of vetoes for the burst and CBC searches. Similar work has been conducted by the LSC in the past~\cite{Nelson05,Nelson04,Cadonati03,Dicredic05}. For burst searches using the third LIGO science run the channel containing 
control signals from the power recycling cavity was used as a veto channel~\cite{Dicredic05}. A discusson
of various data quality flags and veto channels used for burst searches with data from the fourth LIGO science
can be found in Ref.~\cite{bursts4}.


We provide a few examples of how our work helped commissioning efforts and improved data quality during the fifth LIGO science run. 
Although a detailed summary of all the investigations done by the glitch group over almost two years is beyond the scope of this paper, we list some of the most relevant results obtained: 

$-$Near the start of S5, we were able to track down causes of lock loss in the Livingston interferometer
due to ``channel hopping'', when signals intended to drive the amplitude of an auxiliary laser was instead injected in a channel pushing the detector mirrors. These are similar to the example of the calibration glitch shown in
Fig.~\ref{edspecgram}. 
The control and data system was fixed to monitor and prevent this artifact in the rest of the run. 

$-$ We found that many coincident H1-H2 glitches were also coincident with
events in magnetometer and voltage channels (Fig.~\ref{qscanplot}). Our
investigation of several of these events revealed that they coincided
with circuit breaker trips, shorts, and other faults in high-voltage
transmission lines that are connected to power substations near
Hanford. The effect of these power grid events on the interferometer
was consistent with what we expected from coupling of the ambient
magnetic field transients to the permanent magnets on the test masses.
The times of these and other power grid disruptions were flagged to
prevent false alarms.

\begin{figure}
\begin{center}
 \includegraphics[width=0.5\textwidth]{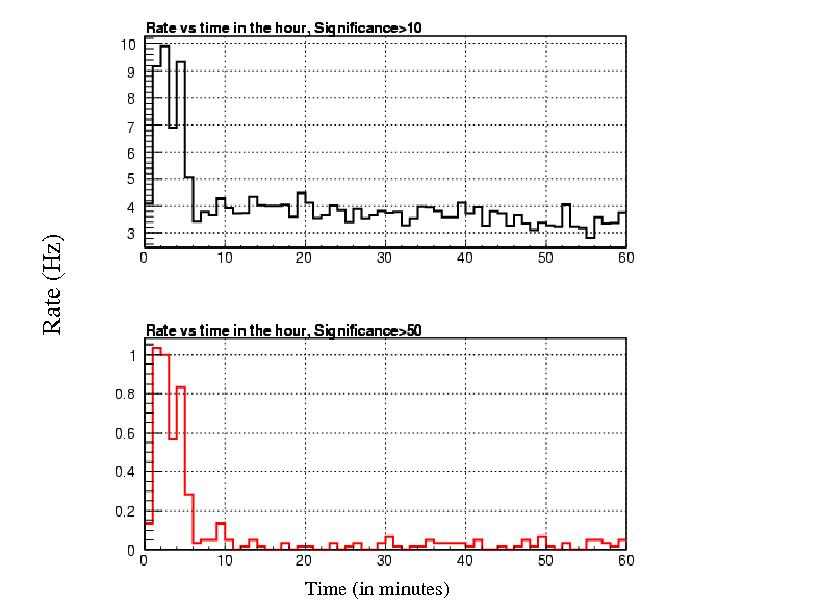}
  \caption{ \label{kwhist} Histogram of the {\tt KleineWelle} L1 event rate within an 
hour over a one day period when the detector experienced hourly  glitches. The top and 
bottom panels shows the rate of  events with {\tt KleineWelle}  significance 
greater than 10 and 50 respectively.}
\end{center}
\end{figure}

$-$Starting from October 2006, L1 experienced periodic glitches near the beginning of the hour,  which were recognized
as due to the digital snapshots of the various 
diagnostic information about the detector which happened once every hour. This was found by looking at the 
histograms of the {\tt KleineWelle} event rate as a function of time within an hour as shown in Fig.~\ref{kwhist}.

$-$ We tracked down the cause of  a few outliers in the gravitational-wave channel to  asymmetric 
response  in the four photodiode signals used in the optical setup. 
Dedicated  monitors  were written following these findings to look 
for similar glitches caused by asymmetric photodiode response throughout S5.
The cause of such glitches was believed to be due to dust along the beam path 
to the photo-diodes and was confirmed  by specalized glitch injections in 
the post S5 period.


$-$ We have also done a classification of data quality flags into four categories with different levels
of severity. These are classified into  Category 1 (which includes data that won't be analyzed), Category 2 (where vetoes
will be applied only in post-processing), Category 3 (which are advisory flags used for detection confidence)
and Category 4 (which are advisory flags used to exert caution in case of a detection candidate). All these
flags will be used for forthcoming burst and CBC papers using S5 data.


\section{Conclusions and future work}
\label{sec:posts5}
Due to the long duration of S5 run, work is still in progress to wrap up all the S5 related glitch 
group efforts.  The most important S5 related task still in progress is the creation of data quality flags 
and this is being done in collaboration with members from the {\sf Data Quality} group. Another 
major effort is to  follow-up possible coincident events
from burst and CBC searches to assess the data quality at the time of the candidate and thus {their statistical significance~\cite{followup}. 
 After S5, there were a few externally induced glitches and environmental injections.
Some work has started  using event visualization tools to characterize these glitches. We are also 
providing guidance to the {\sf Dataset Reduction} group regarding choice of channels and 
sampling rates which need to be archived for the current Astrowatch program  and future LIGO science runs.
Thanks to the systematic effort of the glitch group, many artifacts were identified and the times were flagged, producing better data quality  which allows for better astrophysical results, as well as improved confidence in any candidates that may be identified in the future. Many of the tools developed will be used in future runs for automated identification of the artifacts. The group effort has been very successful, and will likely continue and be improved in the future.

\section*{Acknowledgments}
LIGO was constructed by the California Institute of Technology and  Massachusetts Institute of Technology
with funding from the National Science Foundation and operates under cooperative agreement PHY-0107417. 
We would like to thank the Bonneville Power Administration,
particularly Mike Viles, for providing and interpreting records of
power grid events. We are grateful to  all members of the LIGO Scientific Collaboration for feedback and 
support for our work and especially to all the operators and members from  the commissioning team  who 
tirelessly worked for two long years to ensure a high duty cycle and good sensitivity during the fifth  
LIGO science run. This paper was assigned LIGO document number LIGO-P080016-01-Z.
  
\end{document}